\documentclass[%
superscriptaddress,
longbibliography,
preprint,
amsmath,amssymb,
aps,
prx,
]{revtex4-1}

\usepackage{array}
\usepackage{color}
\usepackage{graphicx}
\usepackage{dcolumn}
\usepackage{bm}
\usepackage{amsmath}
\usepackage{makecell}
\usepackage{multirow}
\usepackage{adjustbox}

\begin{document}

\title{Uncertainty Driven Active Learning of Coarse Grained Free Energy Models}

\author{Blake R. Duschatko}
\email{bduschatko@g.harvard.edu}
\affiliation{John A. Paulson School of Engineering and Applied
Sciences, Harvard University, Cambridge, MA 02138, USA}

\author{Jonathan Vandermause}
\affiliation{John A. Paulson School of Engineering and Applied
Sciences, Harvard University, Cambridge, MA 02138, USA}
\affiliation{Department of Physics, Harvard University, Cambridge, MA 02138, USA}

\author{Nicola Molinari}
\affiliation{John A. Paulson School of Engineering and Applied
Sciences, Harvard University, Cambridge, MA 02138, USA}
\affiliation{Robert Bosch LLC, Watertown, MA 02472, USA}

\author{Boris Kozinsky}
\email{bkoz@seas.harvard.edu}
\affiliation{John A. Paulson School of Engineering and Applied
Sciences, Harvard University, Cambridge, MA 02138, USA}
\affiliation{Robert Bosch LLC, Watertown, MA 02472, USA}

\date{\today}

\begin{abstract}

    Coarse graining techniques play an essential role in accelerating molecular simulations of systems with large length and time scales. Theoretically grounded bottom-up models are appealing due to their thermodynamic consistency with the underlying all-atom models. In this direction, machine learning approaches hold great promise to fitting complex many-body data. However, training models may require collection of large amounts of expensive data. Moreover, quantifying trained model accuracy is challenging, especially in cases of non-trivial free energy configurations, where training data may be sparse. We demonstrate a path towards uncertainty-aware models of coarse grained free energy surfaces. Specifically, we show that principled Bayesian model uncertainty allows for efficient data collection through an on-the-fly active learning framework and open the possibility of adaptive transfer of models across different chemical systems. Uncertainties also characterize models' accuracy of free energy predictions, even when training is performed only on forces. This work helps pave the way towards efficient autonomous training of reliable and uncertainty aware many-body machine learned coarse grain models.
\end{abstract}

\maketitle

\section{Introduction}

Molecular dynamics (MD) has long been used to efficiently study the statistical and kinetic properties of a wide variety of systems by integrating Newton's equations of motion \cite{allen-tildesley, smit-frenkel}. There are, however, substantial limitations on the applicability of molecular dynamics to situations where the behavior at long length and time scales is relevant, such as many biological processes. Fast degrees of freedom in the system prohibit the use of larger integration time steps while creating a more rugged energy landscape that typically slows important structural changes~\cite{martini, protein_models}. Exploring a rough energy landscape with small time steps becomes unfavorable when coarser resolution is of interest. In addition, computing forces and updating the configuration state of every degree of freedom (DOF) at each time step in an all-atom (AA) system can be a computational burden. For a variety of problems, fast motions such as those of hydrogen vibrations do not play a significant role in long length- and time-scale properties, making it unnecessary to track each DOF. 

In the context of biological systems, understanding protein folding pathways and the bulk properties of lipid bilayer membranes often require simulations of physical timescales on the order of microseconds or longer, while also typically modeling tens or hundreds of thousands of atoms~\cite{aa-lipid-bilayers, coiled-peptides}. In the most cutting-edge applications of MD, on the other hand, it is not typically feasible to exceed these two limits simultaneously. Similarly, at the all-atom scale protein-protein interactions, or more generally polymer-polymer interactions, may not depend on all degrees of freedom.  Accounting for the effects of solvents adds additional computational complexity. In such scenarios as described above, a variety of coarse grained (CG) techniques are often used to probe the system at longer time scales and lower spatial resolution~\cite{bonomi_reconstructing_2009, torrie1977nonphysical,barducci_metadynamics_2011,souaille2001extension,protein_models, self-assembly, amber, martini, martini-proteins, martini-carbohydrates,martini-improved, martini-polarizable, oplsua, cabs, ua-alkanes, amber, rosetta, unres, noid_cgPerspective, clementi_solvent, cg_phase_transition, qm_cgmm, dft_qm_cgmm}. 

Of central importance to modeling the thermodynamics of systems with CG approaches is identifying the reduced degrees of freedom, or CG beads, and determining the interactions between them. To this end, two primary approaches are taken. Top-down methods are parametric models tuned to reproduce experimental observations, while bottom-up methods are built upon an underlying all-atom description of the system. Top-down coarse grained force fields are well established with widespread use across different applications \cite{oplsua, martini, martini-proteins, martini-improved, martini-polarizable, martini-carbohydrates,amber, cabs, rosetta, unres, protein_models}, while the development of bottom-up coarse grained models has seen substantial activity within the past decade or so \cite{mscg, mscg-protein, mscg-lst, mscg-threebody, rel_entropy, mscg-liquids, clementi_solvent, clementi_mlcg, deepcg, rafa_temp, clementi_gnn, autoencoders, gap_cg, gdml, gp_cg}. Bottom-up approaches can rigorously reproduce the statistical behavior of the system of interest by targeting the many-body coordinate dependent free energy, often called the potential of mean force (PMF), represented by a reduced set of coordinates \cite{mscg, rel_entropy}. 

Due to the complexity of the PMF, machine learned (ML) forces fields have recently gained traction over empirically parameterized classical potentials. In particular, the success of all-atom machine learned force fields as surrogate models for \textit{ab initio} molecular dynamics \cite{behler_parrinello, behler_symmetryFunctions, gap, chemical_envs, otf, covariant_kernels, nbody_expansion, schnet, flare, flarepp, mappedGP, nequip, allegro, bartok2010gaussian, bartok2013representing, bartok2015gaussian} has resulted in increased interest in applying similar techniques to modeling the PMF \cite{clementi_solvent, clementi_mlcg, deepcg, rafa_temp, clementi_gnn, autoencoders, gap_cg, gdml, gp_cg}. These approaches typically make use of regression over long all-atom trajectories via a multiscale coarse graining / force matching technique that reproduces the all-atom PMF in the limit of sufficient sampling of the canonical ensemble~\cite{mscg}. In addition, machine learning has targeted related problems of choosing an optimal low-resolution representation of atomic systems \cite{autoencoders} and more broadly the problem of reconstructing atomic details from CG models \cite{backmap_bayesian, backmap2, tess_structure_gen}. 
So far, most ML approaches to CG models were based on neural networks (NN), which possess a number of benefits. For example, they serve as highly flexible representations of complex functions that can be trained on large training sets. Kernel-based methods have also been proposed \cite{gap_cg, gdml, gp_cg}. However, all ML-based CG models of the PMF to date lack a measure of principled uncertainty.

The wide array of problems that are faced in CG modeling motivate a new perspective. For instance, in exploratory applications, such as the discovery of structure of large proteins or rapid screening of soft materials, training CG models with a set of relevant configurations may become infeasible. Some configurations may never be sampled due to their rarity or the time it takes for a system to evolve into these states.  For this reason, current approaches that do not use uncertainty metrics rely heavily on substantial \textit{a priori} knowledge of the target behavior of the all-atom system, posing a limitation on their potential for wide spread applicability. 

Moreover, it becomes increasingly difficult to assess the quality of models by estimating the true error on the test set, since this requires long constrained dynamics trajectories to obtain well-converged PMF data. In addition, traditional aggregate metrics such as mean absolute errors of forces may not capture subtle deficiencies in a model. Also, bottom-up CG models are highly dependent on the configuration and chemical make up of the all-atom system used for training, making the transferability of these CG models difficult to anticipate. 

Compared to NNs, kernel-based Bayesian regression methods, such as Gaussian process (GP) regression, provide access to predictive uncertainty and have been demonstrated to efficiently select sufficiently representative training sets via active learning in all-atom settings~\cite{flare, flarepp, nbody_expansion}. One limitation is the increase of the computational cost of the training and inference tasks with the size of the training set. However, in many cases the full GP can be mapped onto an exact model for predicting both the mean and uncertainty with a cost that is independent of the training set size. Application of these recent methods have been demonstrated to simulations of complex heterogeneous systems at record speed and size, reaching 500 billion atoms ~\cite{mappedGP, anders-micron, yu-sic}. 

In this work, we present an active learning regression framework based on principled Bayesian uncertainty inherent to sparse Gaussian processes (SGP's) for autonomous development of coarse grained force field models. First, we demonstrate the ability of Gaussian process regression to learn the coarse grained PMF on-the-fly, thereby reducing the guess work typically needed to construct the training set. In practice, this allows the model to discover unknown configurations that may appear at long timescales by bypassing the more predictable motions of faster degrees of freedom. Second, we show that uncertainty-aware active learning enables the development of more transferable coarse grained models. In particular, we demonstrate how uncertainty, along with the locality, allows models of one molecular system to be "transferred" to a new system by updating the training database. Third, we find that uncertainty allows for a rapid and direct assessment of model robustness and limitations without the need for test set data, accelerating deployment and facilitating automatic refinement.

\section{Results} \label{results}

\subsection{Validity of uncertainty aware on-the-fly learning of coarse grained models}

Bottom-up training of coarse grained models is typically done by regressing the PMF derivatives to time averages of forces, the precise form of which is given by equations (\ref{eq:pmf}) and (\ref{eq:meanforce}). A common approach is to utilize instantaneous forces in a long unconstrained molecular dynamics trajectory to minimize appropriate functionals that reproduce the PMF~\cite{mscg, clementi_mlcg}. This approach requires care in dealing with two implicit timescales in the problem. For one, the simulation times of the fast degrees of freedom must be long enough to ensure that the sampled all-atom configurations are not highly correlated. Also, collecting sufficient training data requires simulating long enough time scales in order to visit a full range of coarse grained configurations. Even so, depending on the shape of the PMF, some regions in CG configuration space separated by barriers may not be visited during the training simulation. Principled quantitative uncertainty provides a rigorous way of identifying configurations that lie within and outside the training set.

\begin{figure*}[h!]
    \centering
    \includegraphics[scale=.025]{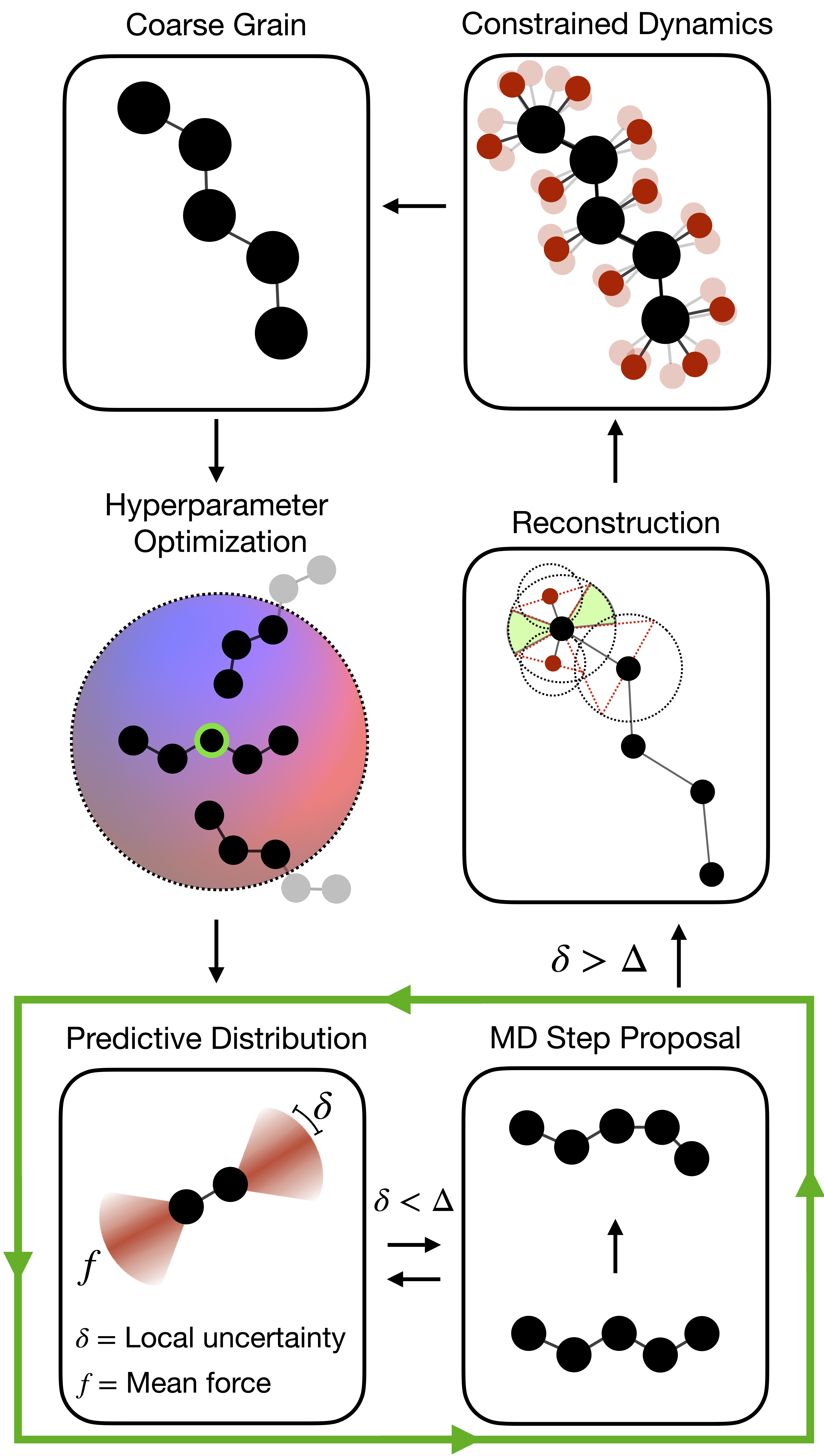}
    \caption{FLARE-CG is an extension of the Fast Learning of Atomic Rare Events software~\cite{flare,flarepp,mappedGP}. Here we schematically demonstrate the workflow of the on-the-fly active learning training loop. An initial all-atom frame is run under constrained dynamics and coarse grained to obtain initial force labels. A select number of sparse environments are randomly added to the training set of the Gaussian process. The construction of model descriptors is performed, and the hyperparameters of the SGP updated by maximizing the log marginal likelihood. The model proposes a molecular dynamics step, along with force and local free energy uncertainties. If all local free energy uncertainties are below a tolerance threshold, the step is accepted. Otherwise, reconstruction is performed in order to collect more constrained dynamics training data. To perform reconstruction, we construct excluded volumes around already placed atomic centers in the system. New atom placements are proposed by randomly drawing an azimuthal and polar angle pair, which are subsequently accepted if they do not lie within regions of overlap between existing excluded volumes.}
    \label{fig:diagram}
\end{figure*}

Here, we introduce a novel on-the-fly CG workflow implemented in the FLARE framework~\cite{flare,flarepp}, depicted in Fig.~\ref{fig:diagram}, that utilizes the predictive Bayesian uncertainty of SGPs. The goal of this FLARE-CG approach is to automate the collection of the PMF labels by deriving a decision threshold of data acquisition from the uncertainty associated with every prediction during a CG MD simulation. This is implemented by directly comparing each local CG environment for which predictions are made with those in the training set using a pre-defined kernel function operating on geometric descriptors of local configuration environments. Specifically, we consider the local environment of CG sites within a defined cutoff radius from the central site. Many-body symmetry preserving descriptors based on the atomic cluster expansion (ACE) are constructed for this purpose, with different hyperparameters allowing for different levels of expressiveness~\cite{ace, flarepp}. 

The workflow begins from an initial coarse grained structural configuration, for which a constrained dynamics trajectory is performed to acquire force labels for training. Subsequently, for each configuration, the predicted models local PMF uncertainties are used to decide whether to evolve the system forward in time. If the uncertainty of the model on the new configuration is above a user-defined threshold, more constrained dynamics data is collected to augment the Gaussian process model database; otherwise, the CG MD step is accepted, and the system evolves forward by a time step. This process is repeated at every step of the CG MD trajectory, forming an autonomous active learning loop. Eventually, after the model no longer makes frequent calls to the all-atom baseline, the trained SGP model's explicit dependence on the training set size can be eliminated by mapping it onto an exactly equivalent and much faster parametric model~\cite{flarepp}.

\begin{figure*}[h!]
    \centering
    \includegraphics[scale=.87]{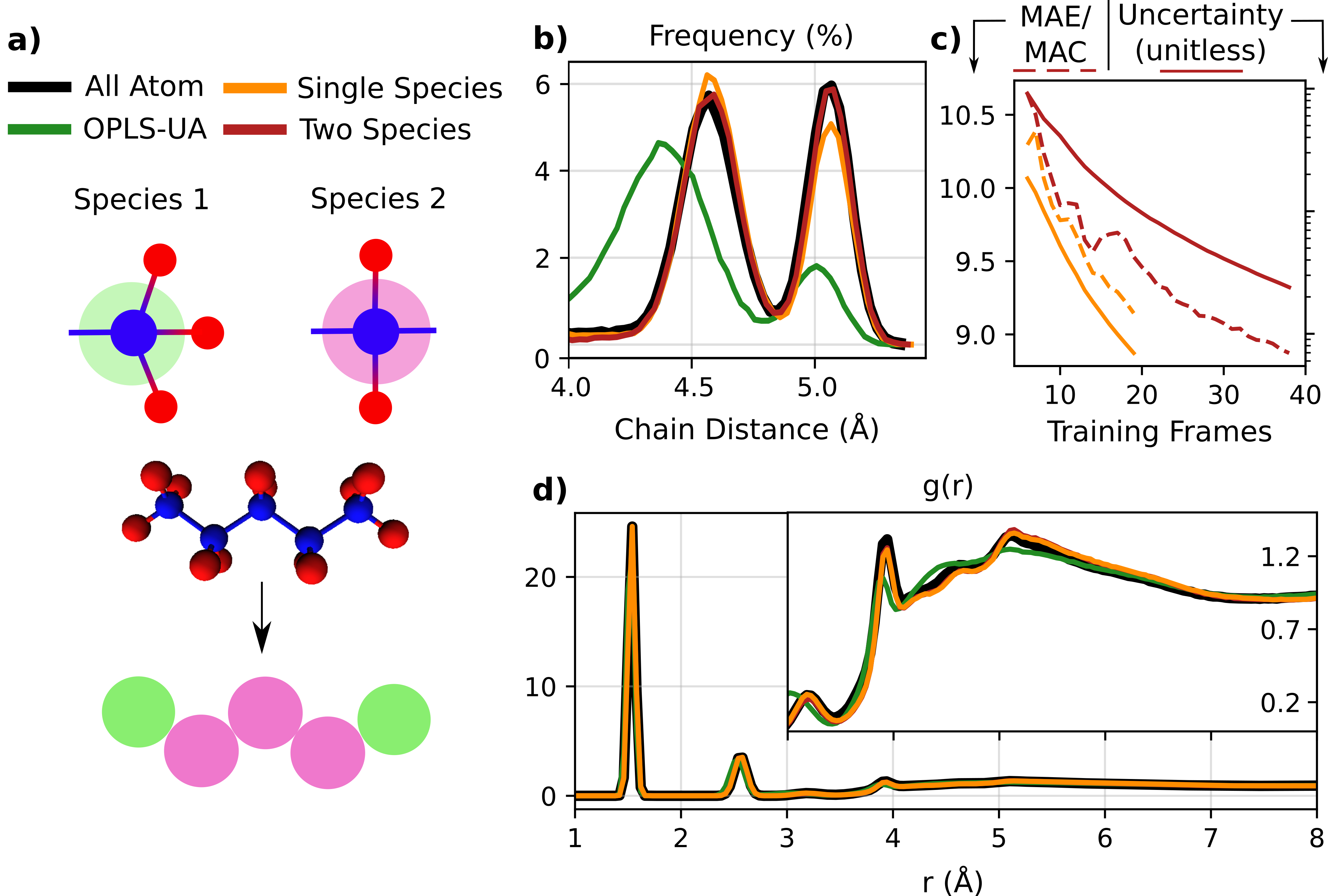}
    \caption{a) atoms in a pentane liquid are mapped directly to their carbon sites, integrating out hydrogen degrees of freedom. In two experiments, the beads are treated as either one or two different species, based on the underlying bond topology. b) the end-to-end chain distance distribution of single and two species models are compared to the all-atom training baseline, as well as a common coarse grained force field for hydrocarbon liquids, OPLS-UA. c) the learning rate for two and single species models is reported by showing the mean absolute force error on a test set as a relative percentage to the mean absolute force component in each test frame. We also report the mean free energy uncertainty of the model on the set set, defined in equation~\ref{eq:uncertainty}, which is a unitless quantity. d) the full carbon-carbon radial distribution function for single species, two species, all atom, and OPLS-UA simulations. The inset shows a zoomed-in view of the long-range structure. }
    \label{fig:learning}
\end{figure*}

During the active learning process, the model hyperparameters are adaptively optimized by maximizing the log-likelihood in response to new training data. By maintaining well calibrated uncertainties, models are then capable of discovering new configurations on their own, circumventing a major problem of potentially missing unknown structures in a model's training set. Further, this methodology allows us to rely on the all-atom models only to remove fast degrees of freedom and obtain CG force training labels, so that the slow degrees of freedom are evolved directly using the learned low-resolution CG surrogate model. We illustrate the performance of SGPs trained on-the-fly for a pentane liquid structure consisting of 70 molecules, with the hydrogen degrees of freedom integrated out (Fig.~\ref{fig:learning}). 

The all-atom to CG mapping sequence is shown schematically in Fig.~\ref{fig:learning}a. Our approach requires a prior definition of CG sites. Although recent works have considered automating the CG site selection process~\cite{autoencoders, equivariant-backmap, backmap_bayesian, backmap_md, backmap1, backmap2}, more work is needed to integrate these approaches with our proposed framework. We instead rely on chemical intuition to choose the CG mapping, where for the pentane system we choose to map the all-atom system to carbon sites. In this example, the hydrogen atoms are integrated out, effectively setting to zero their weight in the mapping function, defined in equation (\ref{eq:mapping}) of Methods.

Existing empirical and ML CG models often label coarse grained species by their underlying all-atom composition and bond topology, suggesting the end carbon atoms of pentane should be treated differently than those on the interior of the chain~\cite{clementi_mlcg, martini}. In particular, CG carbons can be treated as different species types in the descriptor equation (\ref{eq:descriptor}) depending on whether they are at the end of the molecular chain. Alternatively, all carbons can be treated as the same species type, and we can rely on the ML model to correctly learn CG forces from the geometry of the local environment structure. 

We investigate the impact of this choice by exploring the performance of models of pentane with and without explicit labeling of end carbons as a different species in the descriptor. Hydrocarbon systems such as pentane liquid are a convenient test case, as many top-down empirical models are parameterized for them. In the following, we compare the accuracy of our ML CG models, whose all atom baseline is the OPLS force field, to an empirical CG force field with the same CG site mapping, OPLS-UA~\cite{opls, oplsua}. For the ML CG approach, we find in Fig.~\ref{fig:learning}c that the single-species and the two-species models provide similar force accuracy as well as the reproduction of structural properties compared to the all-atom baseline. Such comparisons are valuable because the less complex single-species models have shorter inference times. This arises from the scaling of the dimensions of ACE descriptors and kernel matrices with increasing number of species, associated with more computationally heavy linear algebra computations at inference. We are not claiming, however, that this finding for pentane is a general result expected to hold for other molecular systems.

Similarly, comparing the learning of on-the-fly models for both single- and two-species realizations in Fig.~\ref{fig:learning}c, we find both types of models achieve comparable force accuracy. We note that the predictive uncertainties of each model differ between single- and two- species as a result of having inherently different model complexity. In practice this means that single-species models will tend to make fewer calls to the reference method (constrained all-atom dynamics). In this example, we use a predictive uncertainty value threshold, defined in equation~\ref{eq:uncertainty} of Methods, of 0.02 for the two species case and 0.01 for the single species case to further highlight this contrast in behavior with respect to uncertainty. The final mean absolute force error, as a percentage of the mean absolute force component of the model predictions, on a test set lies around 9\% for both single and two species models. For comparison, we also compute the error in forces predicted by the OPLS-UA force field on the same test set and find an error of 30\%. 

To emphasize that uncertainties are indeed predictive, in Fig.~\ref{fig:learning}c we show the (unitless) mean local PMF uncertainty given by equation~\ref{eq:uncertainty} of Methods in each test set frame. A frame refers to a structure snapshot together with atomic force information. As a function of training set size, the uncertainty correlates directly with the trends followed by the mean absolute force errors on the test set. Even though the quantitative values of uncertainty and force test-set error do not agree, the crucial point is that the same trend behavior persists. 

Fig.~\ref{fig:learning}b and Fig~\ref{fig:learning}d demonstrate that the inter- and intra-molecular properties, respectively, of the CG carbon sites match the behavior of the all-atom carbon atoms with high fidelity. The full carbon-carbon radial distribution functions are well reproduced, as are the end-to-end molecule chain distance. This distance is defined as the linear separation between the two carbons on the ends of each molecule. For completeness, we compare the structural performance of our coarse grained models to an existing empirical (non-ML) model, OPLS-UA~\cite{oplsua}. We find far greater fidelity in the reproduction of structural properties with our SGPs compared to this empirical model, further motivating the use of ML approaches to PMF modeling. 

\subsection{Transferability across molecular systems}

Bottom-up CG models are typically developed to preserve the partition function of the Boltzmann distribution maintained by thermostats in the all-atom MD simulation, thereby ensuring consistency with the AA thermodynamics. However, this implicitly assumes that the model will be specific to both the chemical make up of the all-atom system as well as the thermodynamic state points. Particularly in this setting, where models are specific to a given system, predictive model uncertainty is helpful in quantifying the distance of local CG configurations in the test set from those in the training set. This enables us to systematically adjust and improve CG models.

\begin{figure*}[h!]
    \centering
    \includegraphics[scale=.89]{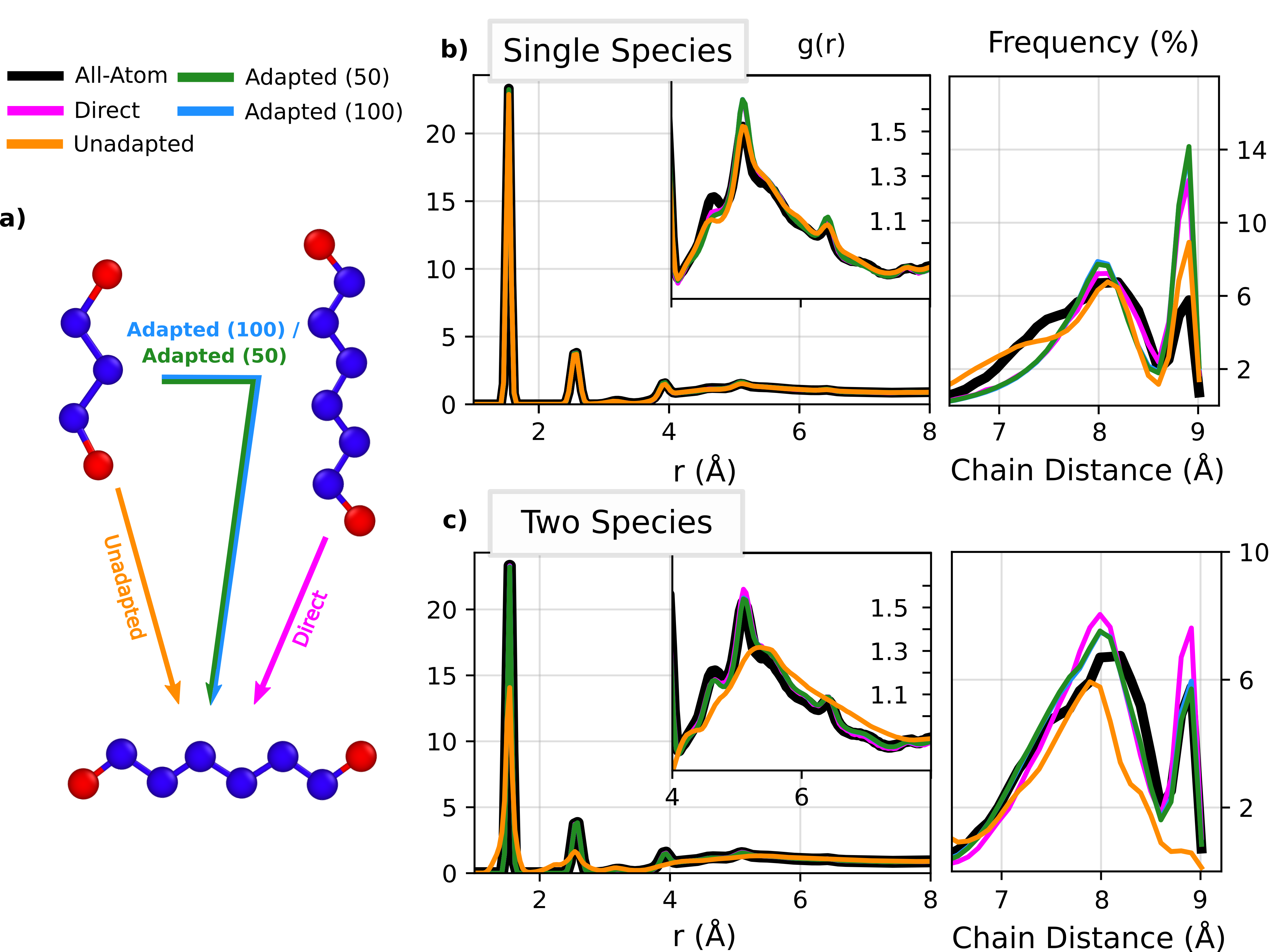}
    \caption{a) a schematic representation of the computational experiment considered. Adapted (50 and 100) pentane models are those that have seen extra octane data, generated on the fly for 50,000 and 100,000 steps, respectively. Unadapted models are pentane models deployed directly on octane with no additional training. b) the carbon-carbon radial distribution function and the end-to-end chain distance frequency are shown for single species models. The inset shows the long-range structure of the liquid c) the carbon-carbon radial distribution function and the end-to-end chain distance frequency are shown for two species models, with the inset showing the long-range structure.}
    \label{fig:transfer}
\end{figure*}

To examine the degree to which SGP CG models trained on one system can "transferred" to another system, we consider several ways in which a model trained on the pentane liquid can be applied and adapted to an octane liquid, as summarized graphically on Fig.~\ref{fig:transfer}a. "Direct" CG models of octane are trained on all-atom octane data on-the-fly using active learning. "Unadapted" CG models are trained on-the-fly on the pentane system for 100,000 time steps and then deployed directly on octane. Here, the SGP evolves the system in CG space over time for 100,000 steps, checking the uncertainty against the user defined threshold at each step, before being deployed on octane. Note that 100,000 does not refer to the number of AA constrained dynamics steps used when collecting more data. The "adapted (50)" and "adapted (100)" models have identical pentane training data, but their training sets are subsequently augmented on-the-fly for 50k and 100k time steps with octane training data, respectively, via the same procedure defined for the pentane on-the-fly loop. We also compare the performance of trained CG pentane models to the underlying AA results for octane. The goal of this experiment is to determine whether or not uncertainty can enable models previously trained on other systems to be extended to a new system more efficiently than starting over and without loss of accuracy. The results of each model type shown in Fig.~\ref{fig:transfer} and Table~\ref{table:transfer} is the average of 40 independently trained models.

We find in Fig.~\ref{fig:transfer}c for the two species models that the inter- and intra-molecular structure of the carbon atoms is reproduced substantially better with the adapted models than their unadapted counterparts. This is further reflected in the end-to-end chain distance RMSE values given in Table~\ref{table:transfer}. Note, however, that the unadapted single-species models shown in Fig.~\ref{fig:transfer}b give higher fidelity radial distribution functions compared to the unadapted two species models while using far less training data on pentane (see Table~\ref{table:transfer}). This result can be attributed to the higher dimensionality of the two-species descriptors, which makes the two-species model more sensitive to changes in local environments. In the single-species case, for example, local environments near the end of the hydrocarbon chains, look highly similar to those of octane. For the two-species case, this is not true as the local environments, within the 4.5 \AA cutoff, of octane carbon sites will almost never contain both ends of the chain, which is likely to happen in pentane. As a result, to a single species kernel, octane looks more similar to pentane and it is able to extrapolate with a limited amount of data.

To highlight the benefits of using an adapted model on systems outside of the training set, we compare the computational cost of AA constrained dynamics reference computations as well as the resulting force errors and structural properties of adapted models compared to training an octane model from scratch (Table~\ref{table:transfer}). Assuming the pentane data is already available, we find that by using an SGP containing data from a chemically similar system, far fewer AA reference calls to the new system are needed compared to starting from scratch. In both the single- and two-species cases, the force accuracy of adapted models with fewer active learning calls to the octane reference AA constrained dynamics method is higher than in direct models trained on octane from scratch which also requests more data in active learning. Additionally, over the course of the 50,000 and 100,000 on-the-fly training steps, we report in Table~\ref{table:transfer} the average number of frames in which the models request more octane data. We see explicitly that in all cases, models starting from pentane request fewer frames of data for the new system. This is a crucial point, as the reduction in required constrained dynamics calls of adapted models significantly reduces the computational cost of model training. We observe that the single- and two-species models display comparable force errors despite their disagreement in the reproduction of chain distance distributions, which we explore further in the next section.

\subsection{Correlating Model Robustness and PMF Uncertainty}


In this section we aim to examine the correlation between the force error and error in structural properties determined by the PMF. Specifically, we focus our attention on the end-to-end chain distance distribution. Physically, the multi-modal distributions that appear in pentane and octane systems arise from the PMF minima at two values of dihedral angles along the chain. For shorter chains, fewer pairs of dihedral angles exist that are energetically favorable, whereas longer chains effectively become more flexible to bending.

To quantify the variation between a model and the AA system in the end-to-end chain distance distribution, we define the population error as 
\begin{equation}\label{eqn:population-ratio}
   100 \times \Bigg|\frac{p_{1}^{cg}/p_2^{cg}}{p_1^{aa}/p_2^{aa}}-1\Bigg|
\end{equation}
where $p_{1,2}$ are the populations (i.e. frequencies of sampling end-to-end chain distance values) of the first and second PMF basins of the AA or CG system. 

With this definition, we can motivate this discussion by considering the relationship between mean force errors and population errors for the octane models discussed in the previous section. The mean force errors of the adapted single-species models are quite similar to the two-species models, while the population error between the single- and two-species approaches differ substantially (see Table.~\ref{table:transfer}). The population error of the adapted (100) single- and two-species models differ by nearly a factor of 10. This result points to a clear disconnect between force and PMF errors. 

\begin{figure*}[h!]
    \centering
    \includegraphics[scale=.85]{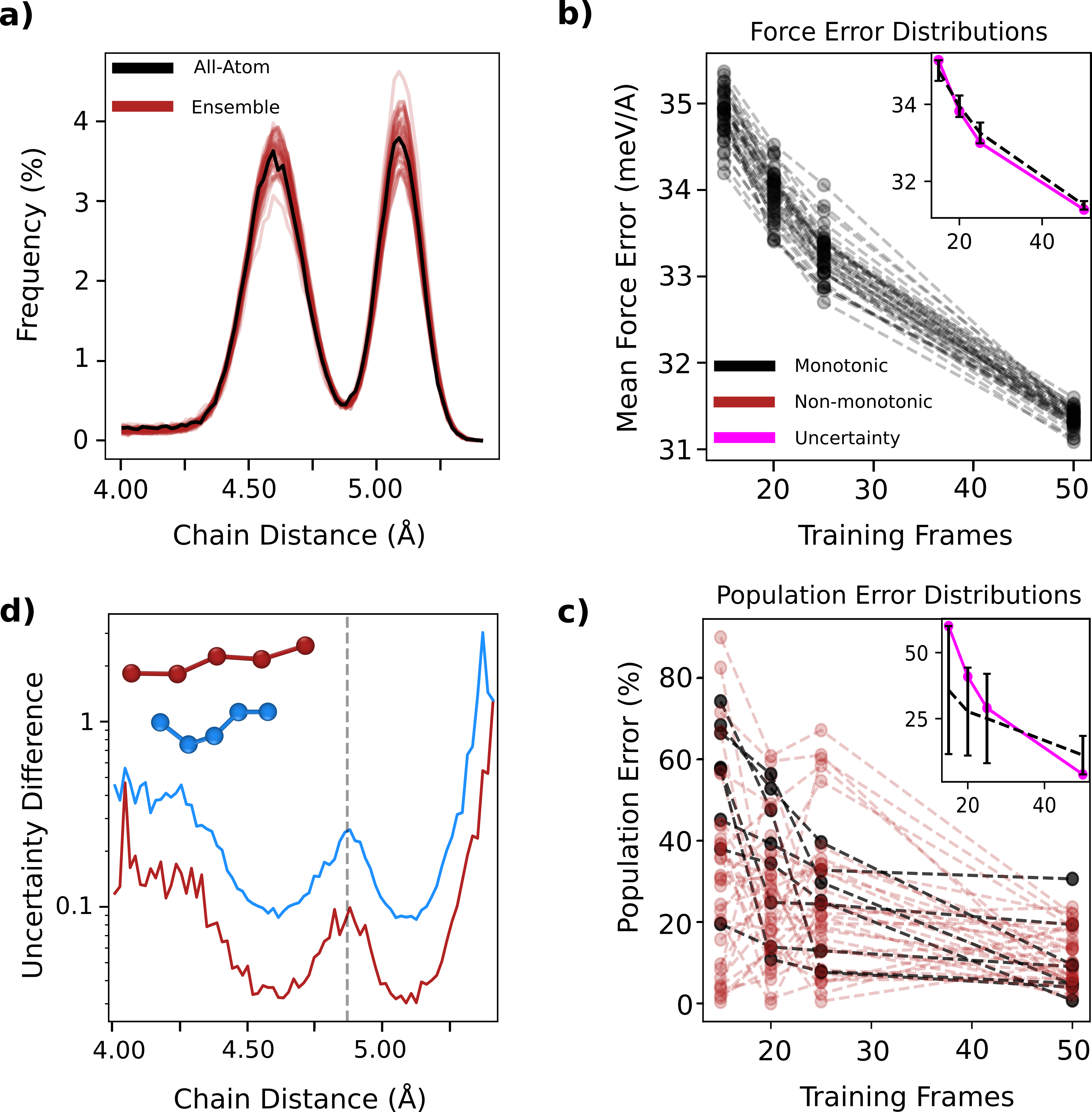}
    \caption{a) the end-to-end chain distance distribution of an ensemble of 40 pentane models, each trained with 50 frames of data b) the distribution of mean absolute force errors and c) the population errors defined in Eqn.~\ref{eqn:molecular1} for the pentane ensemble as a function of training set size. Red lines correspond to non-monotonic trends, while black are monotonic. The insets show the mean and standard deviation of the computed property (black), as well as the uncertainty defined in equation~\ref{eqn:molecular1} averaged over models in the ensemble and over molecules within the transition region (magenta). The uncertainty does not share an axis with the force or population errors, but is simply a unitless quantity as described in the Methods d) the effect of adding different types of data is shown. The carbon sites of a transition state molecule (blue) and unphysically stretched molecule (red) are added as new data to a model having 50 frames of data. The resulting change in molecular uncertainty from the baseline model is shown as a function of chain distance.  }
    \label{fig:chain_distribution}
\end{figure*}

For the remainder of this section, we focus our attention on the case of pentane specifically as it has a much more bimodal distribution in chain distances compared to octane, corresponding to a more challenging energy landscape in which to study population ratio errors. Even longer chains do not display such bimodal distributions at all (see SI Fig. 3). To this end, in Fig.~\ref{fig:chain_distribution}a we show the end-to-end chain distance distributions for a set of 40 pentane models, each trained with subsets of 50 training frames and 50 sparse points per frame, drawn randomly from the same set of AA constrained dynamics data. It is clear that models vary widely in predicting end-to-end pentane chain distance distributions, despite all models exhibiting low errors on forces. Quantitatively, we show in Fig.~\ref{fig:chain_distribution}b the learning curves for the set of 40 models along with their corresponding population errors defined by equation~\ref{eqn:population-ratio}. Black and red lines correspond to monotonically and non-monotonically decreasing quantities, respectively. Each model in the ensemble exhibits monotonic decrease in the force error with more force training data. At the same time, many models show non-monotonic evolution in the population error as a function of the training set size.

These trends can be understood by noting that pentane chain distance distributions are characterized by the difference in PMF values between the two basins. However, in our training, only PMF derivatives (forces) are used, which only implicitly constrain the model's estimate of PMF. As a result, in a test set with minimal representation of transition structures and high sampling of equilibrium ones, we expect that mean force errors will decrease with added data more rapidly than the PMF errors. Thus, improving the fidelity with which these models reproduce distributions of structural properties requires a large amount of force data, especially in the rarely sampled transition region. We note that despite the variability across models in the ensemble, the mean population error as a function of training set size indeed decreases, as does the mean force error. The insets of Fig.~\ref{fig:chain_distribution}b show the mean force error and population error of the ensemble of models, along with the standard deviation represented with error bars. 

The overlap in the distribution of force errors is far smaller than that of the population errors. In each step of the learning process, a model's PMF predictions are less constrained than the force predictions. As a result, there are effectively more learning pathways that quantities arising from the PMF, such as population errors considered here, can take towards a converged value. Because of the substantial overlap in distributions at different stages of training, many of these pathways are not necessarily monotonically decreasing functions.  

One possible solution to minimize this variability would be to include PMF labels in the training set. However, free energies are difficult to compute. In the absence of such PMF labels, however, we argue that by training with force labels alone, the local PMF uncertainties can still be meaningfully interpreted, and that the uncertainties are able to capture useful information regarding the impact force data will have on the model's PMF predictions. 

To connect uncertainty more directly to the performance of observable properties, we explore the local PMF uncertainties of CG environments and their relationship to equation *\ref{eqn:population-ratio}). First, we define the molecular uncertainty as the average local PMF uncertainty on CG sites, $i$, belonging to molecules with an end-to-end chain distance $L$. Formally, this is given by
\begin{equation}\label{eqn:molecular1}
    \sigma(L) = \frac 1N \sqrt{ \sum_{t}\sum_{m_t(L)}\sum_{i\in m_t(L)} \delta \varepsilon_i^2 }
\end{equation}
where $\delta \varepsilon_i^2$ is the local PMF uncertainty of CG site $i$ belonging to a molecule $m$ of length $L$, evaluated on a frame in a test set $t$. The average is performed over all molecules whose end-to-end chain distance lies within a range $L \pm \Delta L$ in the test set. 

In addition to the mean and standard deviation of force errors and population errors shown in Fig.~\ref{fig:chain_distribution}b, we plot the molecular PMF uncertainty in equation~\ref{eqn:molecular1} as a function of the training set size for a single model averaged over molecules within the transition state region. The molecular uncertainty follows the monotonic trend of the true population error averaged over the ensemble of models, while individual models do not necessarily follow such monotonic trends in population errors, as mentioned above. We suggest that, due to the wide variation across models, the local PMF uncertainty for a single model does not necessarily predict the true error of populations on a single model, but rather correlates with the expected average error of an ensemble of models.

We also find that force training labels alone constrain PMF predictions in a meaningful way. In particular, the "usefulness" of the force data that is added is also reflected in the local PMF uncertainties, supporting its use as a metric in on-the-fly training.  To demonstrate this, we examine the molecular uncertainty defined by equation (\ref{eqn:molecular1}) as a function of chain distance for two different scenarios. Starting from a baseline model trained with 50 frames of pentane force data, we compute $\sigma(L)$ on an independent test set. Subsequently, we consider the addition of force labels in a molecule whose chain distance lies within the transition region (\~ 4.8 \AA, model A) compared to the addition of sites in a stretched molecule of chain distance \~ 6.3 \AA, model B. The stretched configuration in model B is highly energetically unfavorable, and we expect that this data would have a less meaningful impact on the local PMF uncertainties of the model within the more well-explored regions of phase space. 

In Fig.~\ref{fig:chain_distribution}c, we plot differences of molecular uncertainties, $\sigma(L) - \sigma_{A,B}(L)$,  between the baseline model and models A and B as a function of chain distance. Indeed, the transition state model shows a sharper decrease in uncertainty overall compared to the stretched model, directly indicating that the local PMF uncertainties are capable of identifying how force data will constrain the corresponding PMF predictions. 

\section{Discussion}

In this work, we have explored the utility of uncertainty aware machine learning models for coarse graining. The principled Bayesian uncertainty measure of Gaussian processes enables a novel on-the-fly active learning scheme for CG models that allows for automating the creation of training sets. A key aspect of the on-the-fly ML CG method is that it overcomes the time- and length-scale limitations of AA models by integrating the AA system in time only over fast degrees of freedom. In addition, by collecting training data only when necessary during MD, we eliminate the need for specifying \textit{a priori} which configurations our model will need to be trained on. Instead, as the algorithm explores configurations, it automatically decides if they are new enough to be added to the training set.

Moreover, the on-the-fly framework enables models to be transferred across molecular systems in a principled way. By the locality assumption inherent in the structural descriptors, for local environments that are similar across systems forces will be predicted to be similar, at a given thermodynamic state. Similarly, environments that differ will result in small kernel values and contribute proportionally less to the prediction of forces on new environments, and the framework will request more data. 

In practice, this approach could be used to develop CG models for common polymer backbones that could then be efficiently adapted to systems with differing functionalization. In this case, only data around new functional groups would be required, as opposed to starting from scratch. This would allow for more rapid development of coarse grained models in materials screening settings. We emphasize that there is a tradeoff in single- and two-species representations for transferable models. In the octane example considered in this work, there is an ease of direct transferability of single-species models without adaption. On the other hand, the achievable accuracy upon the addition of more data of the more descriptive two-species model may be more desirable. In general, it is difficult to anticipate the extent of this complex tradeoff.

The issue of model variance is further addressed herein as arising from limited direct PMF information. By training on time-averaged forces only, obtained from unbiased MD configurations, PMF in transition regions is difficult to capture. What's more, examining force errors alone can seem to suggest that models should always improve with more data. In fact, we show that over an ensemble of models, average force errors indeed decrease monotonically with more data, while PMF errors do not necessarily decrease for each model. As a result, properties arising from PMF values, such as population ratios of stable configurations, can vary significantly and converge slowly in the training process. We find only that the average over a set of models will improve such property predictions. We demonstrate that despite the lack of direct PMF labels in the training set, providing forces still imparts meaningful PMF information to the model.

We note that the molecular local PMF uncertainty plotted in Figs.~\ref{fig:chain_distribution}b,c does not correspond quantitatively to the mean absolute force error or population error. While some work has been done on understanding how to more concretely link uncertainty and performance~\cite{property-uncertainty}, it is not well understood how we can, for example, translate specific values of uncertainty directly to the variance of observed structural properties over an ensemble of models. This remains an open question and demands systematic investigation even for force-fields in all-atom simulations. Moreover, the particular form in which we analyze molecular uncertainty in equation~\ref{eqn:molecular1} is a choice. Other functional forms of molecular uncertainty can be conceptualized that may or may not provide deeper physical insight. This will require careful considerations in future works. 

Finally, we note that a particular challenge in making the proposed method more generally applicable is the reconstruction of all-atom configurations to enable constrained dynamics during the active learning loop. In order to seamlessly go between the all-atom and coarse grained representations, a scheme for recovering lost degrees of freedom must be designed. In many cases this is done with techniques designed for specific systems (as we have utilized in this work), or brute force methods such as a multi-stage compression and expansion of the system box to equilibrate replaced degrees of freedom. Recent works have begun to examine this problem from the perspective of machine learning~\cite{equivariant-backmap, backmap_md, backmap_bayesian, backmap1, backmap2}, but such approaches require pre-selected training sets to learning the CG mapping that would require additional work to reconcile with our active learning scheme. Doing so will enable a much broader study of materials at a variety of resolutions.

\section{Methods} \label{cov}

\subsection{Potential of Mean Force} \label{pmf}

While many approaches exist for designing coarse grained models, top-down approaches lack thermodynamic consistency. A thermodynamically consistent coarse grained model is one for which the Boltzmann distribution of the coarse grained sites is the same as that implied by the underlying all-atom model \cite{mscg}. In principle, all of the thermodynamics of the all-atom system may be recovered from the resulting potential of mean force. 

Suppose we have an all-atom model consisting of $n$ atoms at a set of positions $\boldsymbol{r}^n = \{\boldsymbol{r}_1, \boldsymbol{r}_2,\dots,\boldsymbol{r}_n \}$ and governed by a potential $u(\boldsymbol{r}^n)$. We consider a mapping of the AA system to $N$ coarse grained sites, given by $\boldsymbol{R}^N= \{\boldsymbol{R}_1, \boldsymbol{R}_2,\dots,\boldsymbol{R}_N \}$, of the form
\begin{equation} \label{eq:mapping}
    \boldsymbol{R}_j = \boldsymbol{M}_j(r^n) = \sum_{i=1}^n c_{ij}\boldsymbol{r}_i
\end{equation}
where $j$ denotes the index of the coarse grained site and $i$ the index of each atom. The mapping coefficients $c_{ij}$ must satisfy $\sum_i c_{ij} = 1$ to ensure translation invariance of the resulting model~\cite{mscg}. 

The above information is enough to define the Boltzmann probability distribution of the AA system within the canonical ensemble, so that 
\begin{equation} \label{eq:aa-boltzmann}
    p(\boldsymbol{r}^n) = \frac{\exp(-u(\boldsymbol{r}^n) / k_BT)}{Z_r}
\end{equation}
with Boltzmann constant $k_B$, temperature $T$, and partition function $Z_r$. Let the coarse grained system be governed by its own potential, $U(\boldsymbol{R}^N)$, so that we can define a Boltzmann probability for the coarse grained sites at the same state points as
\begin{equation} \label{eq:cg-boltzmann}
    P(\boldsymbol{R}^N) = \frac{\exp(-U(\boldsymbol{R}^N) / k_BT)}{Z_R}
\end{equation}

For a consistent coarse grained model, we would like the probability of sampling a given coarse grained configuration to be the same as if we had sampled the sites from the AA system. In other words,
\begin{equation}
    P(\boldsymbol{R}^N) = \int p(\boldsymbol{r}^n)\prod_{j=1}^N\delta(\boldsymbol{M}_j(\boldsymbol{r}^n) - \boldsymbol{R}_j) d\boldsymbol{r}^n
\end{equation}
This requirement defines the potential of mean force, $U(\boldsymbol{R}^N)$, as a coordinate-dependent free energy surface in terms of an underlying AA model. In particular, 
\begin{equation} \label{eq:pmf}
    \exp(-U(\boldsymbol{R}^N) / k_BT) \propto \int p(\boldsymbol{r}^n)\prod_{j=1}^N\delta(\boldsymbol{M}_j(\boldsymbol{r}^n) - \boldsymbol{R}_j) d\boldsymbol{r}^n
\end{equation}
The PMF allows us to define a force on each coarse grained site as a gradient of the free energy. This force captures not only energetic effects, but also the entropic contributions to the free energy arising from the degrees of freedom being integrated out.

Formally, for mappings that take a group of atoms to their center of mass, and for which no atom belongs to more than one coarse grained site, the mean force on each site is \cite{mscg}
\begin{equation} \label{eq:meanforce}
    \boldsymbol{F}_j (\boldsymbol{R}^N) = \Big\langle \sum_{i\in S_j} \boldsymbol{f}_i \Big\rangle_{\boldsymbol{R}^N}
\end{equation}
where $S_j$ is the set of atoms involved in the mapping to site $j$, $f_i$ is the atomistic force on atom $i$, and the average is a weighted ensemble average defined as
\begin{equation}\label{eq:ensemble-average}
    \langle g(\boldsymbol{r}^n) \rangle_{\boldsymbol{R}^N} = \frac{\int g(\boldsymbol{r}^n) e^{-u(\boldsymbol{r}^n) / k_BT}\prod_{j=1}^N\delta(\boldsymbol{M}_j(\boldsymbol{r}^n) - \boldsymbol{R}_j)d\boldsymbol{r}^n }{\int e^{-u(\boldsymbol{r}^n) / k_BT}\prod_{j=1}^N\delta(\boldsymbol{M}_j(\boldsymbol{r}^n) - \boldsymbol{R}_j)d\boldsymbol{r}^n }
\end{equation}
for an arbitrary function $g$ of the atomistic coordinates. Note that it is not a requirement that every atom have a non-zero weight in at least one mapping to a coarse grained site. In the hydrocarbon systems we consider, the hydrogen atoms have corresponding coefficients $c_{ij}=0$ for all CG sites, $j$. 

To model the PMF independent of atomic details, we must define a functional approximation to $U(\boldsymbol{R}^N)$ as well as how the parameters of this function will be fit. In the following section, we describe the sparse Gaussian process model used to approximate the PMF, and the mean force labels in equation (\ref{eq:meanforce}) can be estimated using constrained molecular dynamics. 

\subsection{All-Atom Reconstruction} \label{reconstruction}

In order to go from the coarse grained resolution back to an AA representation, we must define a method for recovering AA degrees of freedom. For the n-alkane systems studied in this work, we use the idea of excluded volumes. In all cases, we have chosen to map AA structures onto a carbon site based structure representation, leaving only hydrogen atoms to be reconstructed.

A schematic of this reconstruction scheme is provided in Fig.~\ref{fig:diagram}. Initially, excluded volumes of radius $R_C$ are placed around each CG carbon site. For each site, we define a procedure for reconstructing all hydrogen atoms that are bonded to the central site. We wish to place a hydrogen atom a distance $R_{C-H}$ away from its bonded carbon, which requires a polar and azimuthal angle to define a complete coordinate. A random pair of angles is drawn, and it is determined whether or not the resulting point lies within a restricted region. 

Such restricted regions are defined in terms of overlapping excluded volumes. In particular, a conical region is constructed whose central axis is defined by the vector from the central carbon to the neighboring atom creating an excluded volume overlap. The angular extent of the conical region is determined by the angle between the central axis and the lines connecting the central carbon to the circular intersection of the edges of the excluded environments. 

Should the angle between the vector from the carbon to the new hydrogen site and the central axis vector of any restricted region be less than that regions angular extent, a new pair of polar and azimuthal angles is drawn for the hydrogen until a position that does not lie in an overlapping region is found. This process is iterated for each hydrogen atom bonded to the carbon site before proceeding to the next carbon atom. Note that in general, $R_{C-H}$ and $R_C$ need not be equal, and in our implementation, $R_{C-H}$ is taken to be the experimental value of 1.118 \AA. 

In addition, for each hydrogen atom placed, a new excluded volume is introduced into the system centered at this location, with a radius $R_H$. In this way, the presence of reconstructed degrees of freedom can influence subsequent placements, allowing for easy avoidance of non-physical AA reconstructions. We find stable reconstruction results when setting $R_C$ and $R_H$ to 1.6 \AA~and 1.1 \AA, respectively. These quantities are comparable with the experimental C-H and C-C bond lengths of 1.118 \AA~and 1.531 \AA.

In order to reduce the time spent on sampling, we place a restriction on our reconstruction algorithm such that if an acceptable sample is not drawn in a predefined amount of time, the excluded volume radii are reduced by 10\%. However, we rarely find this to be necessary. 

Once all hydrogen atoms have been reconstructed, energy minimization (see SI.1) is performed, after which hydrogen velocities are drawn at random from a Boltzmann distribution at 250K and equilibrated with a 0.5 fs timestep for 20,000 steps.

\subsection{Sparse Gaussian Processes} \label{sgp}

Bottom-up coarse graining models often require an accurate description of many-body interactions. Here, we use sparse Gaussian processes to do so. We outline the methodology here, with more detail available in Ref.~\cite{flarepp}. 

\subsubsection{Descriptors of Local Free Energy}

We assume that the PMF can be modeled with a purely local function. In particular, given a set of coarse grained coordinates $\boldsymbol{R}^N= \{\boldsymbol{R}_1, \boldsymbol{R}_2,\dots,\boldsymbol{R}_N \}$ with chemical-type identities $\{s_1,s_2,\dots,s_N\}$, the PMF is given by a sum of local free energy contributions from the CG sites in the form
\begin{equation}\label{eq:energy-decomp}
    U(\boldsymbol{R}_1, \boldsymbol{R}_2,\dots,\boldsymbol{R}_N; s_1,s_2,\dots,s_N) = \sum_i^N \varepsilon(s_i, \rho_i)
\end{equation}
where $\rho_i$ is some description of the local environment of CG site $i$. In particular, the environment is the set of distance vectors between neighboring sites, such that $j \neq i$, within a cutoff radius of $r_{\text{cut}}^{(s_i,s_j)}$ that can, in general, be species dependent. Formally, 
\begin{equation}
    \rho_i = \{ (s_j,\vec{r}_{ij}) \text{ }| r_{ij} < \text{ } r_{\text{cut}}^{(s_i,s_j)} \}
\end{equation}

Following the Atomic Cluster Expansion approach introduced by Drautz~\cite{ace}, the local environment of a site can be projected onto single-particle basis functions
\begin{equation}\label{eq:basis}
    \langle \phi_{n\ell m} | \rho_i \rangle = \phi_{n\ell m}(\boldsymbol{r}_{ij})
\end{equation}
where we choose for $\phi$ a decomposition into radial and spherical harmonic components with cutoff function, $c$,
\begin{equation}\label{eq:basis-form}
    \phi_{n\ell m}(\boldsymbol{r}_{ij}) = R_n\left(\frac{r_{ij}}{r_{\text{cut}}^{(s_i,s_j)}}\right) \cdot Y_{\ell m}(\hat{\boldsymbol{r}}_{ij}) \cdot c(r_{ij}, r_{\text{cut}}^{(s_i,s_j)})
\end{equation}
In this work we use Chebyshev polynomials for the radial basis and spherical harmonics for the angular basis. A covariant tensor can be constructed by summing the basis functions over sites within the local environment as
\begin{equation}
    c_{isn\ell m} = \sum_{j \in \rho_i} \delta_{s,s_j}\phi_{n\ell m}(\boldsymbol{r}_{ij})
\end{equation}
Utilizing the sum rule of the spherical harmonic functions, a rotationally invariant descriptor, $\boldsymbol{d}_i$, is
\begin{equation}\label{eq:descriptor}
    d_{is_1s_2n_1n_2\ell} = \sum_{m=-\ell}^\ell c_{is_1n_1\ell m}c_{is_2n_2\ell m}
\end{equation}

The parameters $n$, $\ell$, and $r_{\text{cut}}$ are hyperparameters that we manually specify. To do so, we maximize the marginal log-likelihood (see SI.2), and use the values $n=12$, $\ell=5$, and $r_{\text{cut}}=4.5$ \AA, independent of chemical type. 

\subsubsection{Model Predictions}

To complete the description of the sparse Gaussian process, we must define a kernel that compares the local environment descriptors. As in Ref.~\cite{flarepp}, we choose a kernel that resembles the smooth overlap of atomic potentials (SOAP) kernel~\cite{bartok2010gaussian, bartok2013representing, bartok2015gaussian} and takes the form
\begin{equation}\label{eq:kernel}
    k(\boldsymbol{d}_i,\boldsymbol{d}_s) = \sigma^2 \Bigg(\frac{\boldsymbol{d}_1 \cdot \boldsymbol{d}_2}{d_1d_2} \Bigg)^\xi
\end{equation}
Here, the hyperparameter, $\sigma$, is optimized by maximizing the marginal log-likelihood during on-the-fly training, and $\xi$ is a chosen parameter that can be used to increase the body-order of the model. 

For a set of sparse coarse grained environments, $S$, that is a subset of a larger training set, $F$, a prediction of the local free energy on a new environment $\rho_i$ can be cast as a sum over the sparse points:
\begin{equation}
    \varepsilon(\rho_i) = \sum_{s\in S}^{N_s} k(\boldsymbol{d}_i,\boldsymbol{d}_s)\alpha_s
\end{equation}
where 
\begin{equation}
    \boldsymbol{\alpha} = (\sigma_n^{-1}\boldsymbol{K}_{SF}\boldsymbol{K}_{FS} + \boldsymbol{K}_{SS} )^{-1}\boldsymbol{K}_{SF}\boldsymbol{y}
\end{equation}
where $\boldsymbol{y}$ is the vector of training force labels, $\boldsymbol{K}_{SF}$ the matrix of kernel values between the sparse set and training set, and $\boldsymbol{K}_{SS}$ the matrix of kernel values between points in the sparse set alone. The noise hyperparemter, $\sigma_n$, quantifies the inherent uncertainty present in the training labels and is another hyperparameter that is tuned via maximizing the marginal log-likelihood.

Such mean predictions with the sparse Gaussian process also allow for posterior predictive distribution variances. For SGP's, computing the variance requires approximate methods, where we choose in this work to use the Deterministic Training Conditional (DTC) approximation~\cite{sgp}. As in Ref.~\cite{flarepp}, we use a further simplified form that is the predictive variance of a fictitious Gaussian process trained on local free energies of the sparse environments alone. The resulting uncertainty on local free energies is 
\begin{equation} \label{eq:uncertainty}
    \tilde{V}(\varepsilon) = \frac{k_{\varepsilon\varepsilon} - \boldsymbol{k}_{\varepsilon S}\boldsymbol{K}^{-1}_{SS}\boldsymbol{k}_{S\varepsilon}}{\sigma^2}
\end{equation}
Lying between 0 and 1, this form gives us a unitless measure in defining uncertainty thresholds during on-the-fly training. We find a relative tolerance of 0.02 to perform well. This is found to be stable from our empirical observations and is consistent with similar values used in Ref.~\cite{flarepp}

The resulting models, following the on-the-fly training trajectory, can be simplified such that the summation over sparse points in the predictive distribution can be computed once and used for all future predictions~\cite{flarepp}. This simplification allows for efficient inference upon deployment of the SGPs.

\subsection{Computational Details}

The LAMMPS package~\cite{lammps} was used for all production simulations. The OPLS-AA~\cite{opls} and OPLS-UA~\cite{oplsua} force fields were used in addition to the SGP models. The on-the-fly MD active learning loop was performed using the Atomic Simulation Environment (ASE)~\cite{ase}, and mapped SGP potentials were used in a custom implementation of a LAMMPS pair-style, available as an executable in the FLARE repository~\cite{flare,flarepp}. We have adapted the on-the-fly framework to coarse grained applications, and the software package, FLARE CG, is available upon request. The specific version of the FLARE code used in the production of these results is also available upon request. Parsing of LAMMPS output files for RDFs was done primarily through Ovito~\cite{ovito}. Parameters for all simulations are provided in the Supplementary Information.  

\section{Data availability}
\noindent
All input and output files from the FLARE training and LAMMPS simulations are available upon request.

\section{Code availability}
\noindent
The code used to perform coarse graining and the integration with FLARE will be made public on github. The version used for the data collected herein is otherwise available upon request. 

\section{Acknowledgements}

The authors thank Yu Xie, Lixin Sun, Cameron Owen, Anders Johansson, Noah Bice, and Benjamin Jensen for helpful discussions. This work was supported by a NASA Space Technology Graduate Research Opportunity, by the NSF through the Harvard University Materials Research Science and Engineering Center Grant No. DMR-2011754, and by a Multidisciplinary University Research Initiative sponsored by the Office of Naval Research, under Grant N00014-20-1-2418. All computational experiments utilized the resources provided and maintained by Harvard FAS Research Computing.

\section{Competing interests}
\noindent
The authors declare no competing financial or non-financial interests.

\section{Author contributions}
\noindent
B. R. D. developed the coarse graining extension code to FLARE and performed all computational experiments. J. V. led the development of the all-atom FLARE code and helped with its adaptation and with interpretation of trained models. N. M. assisted in assessing molecular dynamics results and the physical interpretations of the data, as well as contributed to figure design. B. K. supervised the work and contributed to the theoretical developments. B. R. D. wrote the manuscript, and all authors contributed to manuscript preparation.

\bibliography{cg.bib}

\section{Figure Legends}

\setlength{\tabcolsep}{5ex}
\begin{table}[h!]
\centering
\begin{adjustbox}{max width=\textwidth}
\def\arraystretch{1.5}
\begin{tabular}{|c|c|c|c|c|c|}
\hline
& GP Model & \shortstack{Pentane \\ Frames} & \shortstack{Octane \\ Frames} & \shortstack{Force \\ MAE/MAC (\%)} & Population Error (\%) \\ \hline
\multirow{4}{*}{Single Species} & Unadapted Pentane & 11.45 & 0 & 15.29 & 34.98  \\ \cline{2-6} & Adapted (50) Pentane & 11.45 & 1.77 & 10.10 & 53.71  \\ \cline{2-6} & Adapted (100) Pentane & 11.45 & 2.02 & 10.10 & 52.09  \\ \cline{2-6} & Direct & 0 & 12.1 & 10.18 & 49.72 \\ \hline
\multirow{4}{*}{Two Species} & Unadapted Pentane & 42.42 & 0 & 78.18 & 597.05  \\ \cline{2-6}  &  Adapted (50) Pentane & 42.42 & 15.26 & 10.14 & 9.84 \\ \cline{2-6} & Adapted (100) Pentane & 42.42 & 17.52 & 10.04 & 5.85  \\ \cline{2-6} & Direct & 0 & 54.73 & 10.20 & 13.86 \\  \hline
\end{tabular}
\end{adjustbox}
\caption{The average number of all-atom LAMMPS calls over a set of 40 models, as well as the average total number of training frames in the GP, is reported for both single- and two-species models. In addition, we report the mean absolute force error on a test set, averaged over 10 models, as a percentage of the mean absolute force component (MAC) in each test frame, as well as the population error of the averaged chain distance distribution from the all-atom ground truth, defined in equation~\ref{eqn:population-ratio}.  }
\label{table:transfer}
\end{table}

\end{document}